\documentclass[twocolumn, prl, superscriptaddress,notitlepage]{revtex4-1}
\usepackage{graphicx}
\usepackage{physics}
\usepackage{epsfig}
\usepackage{color}
\usepackage{xspace}
\usepackage{array}
\usepackage{amsmath}
\usepackage{amsfonts}
\usepackage{ulem}
\usepackage[dvipsnames]{xcolor}
\usepackage{colortbl}
\usepackage{bm}
\usepackage[breaklinks,colorlinks,linkcolor=blue,citecolor=blue,urlcolor=blue]{hyperref}

\begin{document}
\title{Dual Phonon Universe in a Spinor Bose Gas}

 \author{Jiabin Wang}
 \affiliation{Institute of Theoretical Physics, State Key Laboratory of Quantum Optics and Quantum Optics Devices, Shanxi University, Taiyuan 030006, China}

\author{Yunfei Xue}
\affiliation{MOE Key Laboratory for Nonequilibrium Synthesis and Modulation of Condensed Matter, Shaanxi Province Key Laboratory of Quantum
Information and Quantum Optoelectronic Devices, School of Physics, Xi'an Jiaotong University, Xi'an 710049, China}

\author{Li Chen}
 \email{lchen@sxu.edu.cn}
\affiliation{Institute of Theoretical Physics, State Key Laboratory of Quantum Optics and Quantum Optics Devices, Shanxi University, Taiyuan 030006, China}

 \author{Ren Zhang}
 \email{renzhang@xjtu.edu.cn}
\affiliation{MOE Key Laboratory for Nonequilibrium Synthesis and Modulation of Condensed Matter, Shaanxi Province Key Laboratory of Quantum
Information and Quantum Optoelectronic Devices, School of Physics, Xi'an Jiaotong University, Xi'an 710049, China}
\affiliation{Hefei National Laboratory, Hefei, 230088, China}
\date{\today}

\begin{abstract}
We show the possibility of simulating a dual universe in a pseudospin-1/2 Bose-Einstein condensate (BEC), wherein two phononic modes experience distinct curved spacetime metrics.
Through ramping the interspecies interaction of the BEC, we observe that one universe expands, and in the mean time, the other contracts. These findings can be directly verified in existing cold-atom experimental setups by detecting the correlators of the scaled density contrast (SDC): 
The sizes of the expanded/contracted universes can be deduced from the SDC's oscillating frequencies, while the expanding/contracting histories are reflected in the SDC's amplitude. Our study demonstrates the great potential of spinor BECs for exploring multiverse-related physics in laboratory settings.

\end{abstract}

\maketitle

{\it Introduction-.}
The length and energy scales of cosmology and artificial systems in laboratories (such as cold atoms, classical and quantum optics, and classical water tank) typically exhibit tens of orders of magnitude difference. This has hindered the studies of cosmology in laboratories for a long time. Recently, the ``tabletop cosmology'' is becoming one of the frontiers of the quantum simulation, which benefits from the analogies between these seemingly very distinct systems \cite{Tabletopcos1,Tabletopcos2,Tabletopcos3,Tabletopcos4}.

Atomic
Bose-Einstein condensates (BEC) are believed to be ideal candidates for ``tabletop cosmology''. The suitability stems from the 
relativistic
nature of low-energy phonon excitations in superfluids. Specifically, the low-energy phononic field can mimic the 
relativistic 
matter field, with the background geometric metric being determined by the atomic density distribution and the inter-atomic interaction. In such a framework, the BEC's phonon velocity plays the role 
of
the speed of light in real spacetime, and 
the concept of ``sonic black holes'' has been raised. Near the event horizon of the sonic black hole, such cosmological phenomena as the Hawking radiation \cite{Hawking1,Hawking2,Hawking3,Hawking4}, the Unruh effect \cite{unruh1,Unruh2,Unruh3}, and the quasi-normal mode \cite{quasi-normal_1,quasi-normal_2,quasi-normal_3} have been studied. It is also notable that, the sonic black hole has also been realized in the classical system, such as water \cite{water1,water2,water3,water4}.

The simulation
of curved spacetime benefits from the high controllability of BECs. The atomic density distribution can be precisely tailored through configurable trapping potentials. The inter-atomic interaction 
can be finely tuned via Feshbach resonances \cite{Feshbach1,Feshbach2}. 
For instance, by periodically modulating the interaction of caesium atoms, an Unruh-like matter wave radiation was observed \cite{Unruh-like}. 
Slowly ramping the scattering length \cite{Curved1,Curved2,Curved3,isotropical_expansion} or the radius of the potential \cite{apidly,expanding-contracting} enables the simulation of  the Universe expansion or contraction.

\begin{figure}[t]
	\centering
	\includegraphics[width=0.45\textwidth]{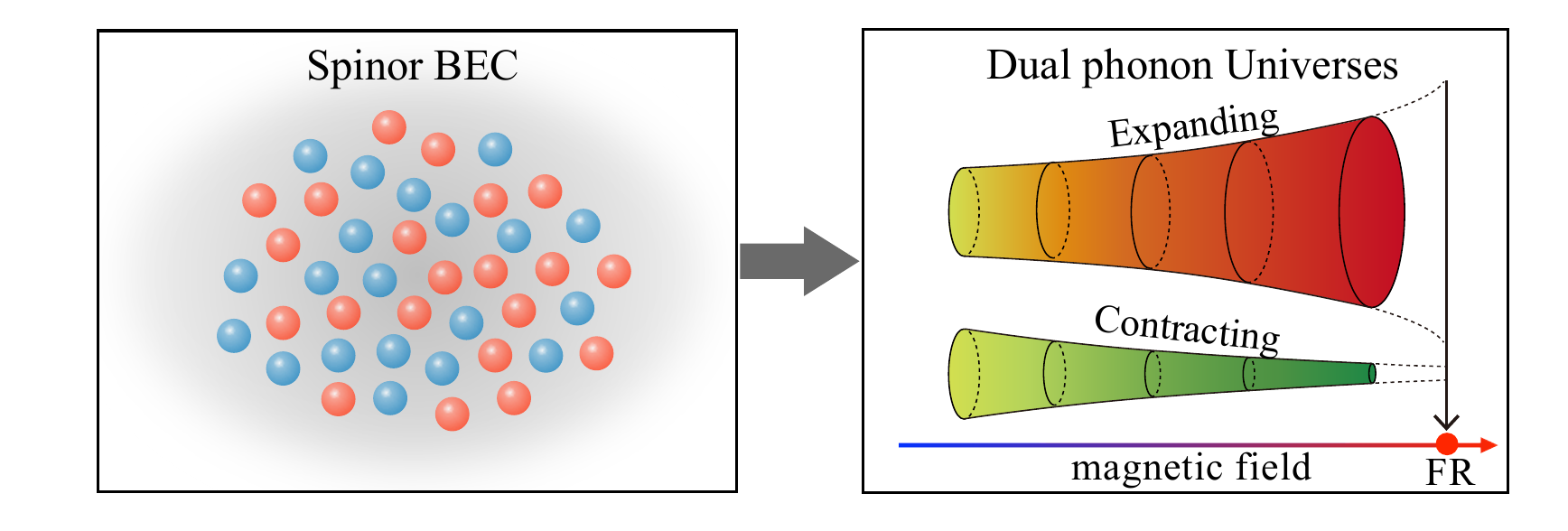}
	\caption{Schematic illustration of the dual phonon universe. Blue and red balls represent the pseudospin-1/2 BEC, in which there exist two modes of phonons, respectively corresponding to the excitation of total and relative fields of the BEC. By ramping the inter-component interaction via magnetic field, one universe expands, while the other contracts.}
	\label{Fig1}
\end{figure}

In addition to the spatial distribution and the atomic interaction, BECs possess another controllable degree of freedom --- the pseudospin. Similar to the spin of elementary particles, pseudospin corresponds to the internal states of atoms. Such BECs with internal states are called spinor BECs, which have played an extremely important role in quantum simulation \cite{simulation1,simulation2,simulation3}, computation \cite{computer1,computer2} and metrology \cite{metrology1,metrology2} over the past decades. However, the pseudospin degree of freedom has been frozen in the experimental simulation of cosmology as of now \cite{Curved1,Curved2,Curved3,apidly,expanding-contracting}. Therefore, the influence of the pseudospin degree of freedom on ``tabletop cosmology'' remains an open question.

In this work, we show a pseudospin-1/2 BEC can simulate a dual universe, with one universe expanding while the other contracting, as schematically illustrated in Fig.~\ref{Fig1}. The expansion/contraction behavior can be well controlled by the inter-species interaction, which is a unique parameter for spinor BECs. During the dynamics, bosons will be emerged from vacuum. 
These phenomena can be observed in existing cold-atom experimental platforms, via detecting the correlators of the density contrast. 
To our best knowledge, parallel universes, as a model for inflation in the early Big Bang, still remain purely theoretical \cite{Big_Bang1,Big_Bang2,Big_Bang3,Big_Bang4,Big_Bang5}. 
Our research suggests a promising route for the laboratory study of multiverse-related physics using spinor BECs.

{\it Dual Phonon Universe-.} We consider a pseudospin-1/2 BEC (labeled by $a$ and $b$) in (2+1)D spacetime. The action is written as 
\begin{equation}
\label{eq:action}
\begin{aligned}
{\cal S}=\sum_{i=a,b}\int dt d^{2}{\bf x}\left[i\hbar \phi_{i}^{*}\partial_{t}\phi_{i}-\frac{\hbar^{2}}{2m}|{\bm \nabla}\phi_{i}|^{2}\right.\\
\left.-V|\phi_{i}|^{2}-\frac{U_{ii}}{2}(\phi_{i}^{*}\phi_{i})^{2}\right]-U_{ab}|\phi_{a}|^{2}|\phi_{b}|^{2},
\end{aligned}
\end{equation}
where $\phi_{a(b)}$ represents the bosonic field of the component-$a(b)$; $V$ denotes the configurable trapping potential; $m$ is the atomic mass; $U_{ii}$ and $U_{ab}$ are the intra- and inter-species interaction, respectively, and both of them can be controlled by Feshbach resonance. The complex fields can be factorized as $\phi_{a(b)}=\sqrt{n_{a(b)}}e^{i\theta_{a(b)}}$, where $n_{a(b)}$ and $\theta_{a(b)}$ are the density and phase fields, respectively. Based on the action $\cal S$, we obtain the equation of motions (EoM) {[see Supplemental Materials (SM) \cite{SM} for details]}
\begin{align}
\label{hydrody}
\hbar\partial_{t}\theta_{i}=&\frac{\hbar^{2}}{2m}\left[\frac{\nabla^{2} n_{i}}{4n_{i}}+\frac{1}{2}{\bm \nabla}\cdot\left(n_{i}^{-1}{\bm \nabla} n_{i}^{-1}\right)-\nabla^{2}\theta_{i}\right]-V\nonumber\\
&-U_{ii}n_{i}-U_{ab}n_{j},\quad i,j=\{a,b\},j\neq i;\\
\label{contunuity}
\partial_{t}n_{i}+&\frac{\hbar}{m}{\bm \nabla}\cdot(n_{i}{\bm \nabla} \theta_{i})=0,\quad i=\{a,b\},
\end{align}
which are hydrodynamic equations of the BEC. 
The inter-species interaction $U_{ab}$ couples the two components.
When $U_{ab}=0$, hydrodynamic equations decouple into two isolated EoM of scalar fields. 

To characterize the fluctuations, we recast the density and phase fields as $n_{i}=\bar{n}_{i}+\delta n_{i}$ and $\theta_{i}=\bar{\theta}_{i}+\delta \theta_{i}$, where $\bar{n}_{i}$ and $\bar{\theta}_{i}$ are the steady solutions of the hydrodynamic equations. 
$\delta n_{i}$ and $\delta \theta_{i}$ denote the fluctuation fields. By substituting the expressions into Eqs.~(\ref{hydrody}) and (\ref{contunuity}), and keeping the linear order of fluctuations, one obtains
\begin{align}
\label{phaseflu}
-\hbar\partial_{t}\delta \theta_{i}&=\hbar {\bf v}_{i}\cdot{\bm \nabla} \delta\theta_{i}+U_{ii}\delta n_{i}+U_{ab}\delta n_{j},\\
\label{densityflu}
\partial_{t}\delta n_{i}&=-{\bm \nabla}\cdot\left[(\delta n_{i}){\bf v}_{i}+\frac{\hbar}{m}\bar{n}_{i}{\bm \nabla} \delta\theta_{i}\right],
\end{align}
where ${\bf v}_{i}=(\hbar{\bm \nabla}{\bar \theta}_{i})/m$ is the velocity field of the superfluid \cite{SM}. Solving Eq.~(\ref{phaseflu}) yields
\begin{align}
\label{soldeltan}
\delta n_{a}=-\frac{\hbar U_{bb}}{U_{aa}U_{bb}-U^{2}_{ab}}\left[\tilde{\nabla}_{a}\delta\theta_{a}-\frac{U_{ab}}{U_{bb}}\tilde{\nabla}_{a}\delta\theta_{b}\right],
\end{align}
where $\tilde{\nabla}_{a}\equiv \partial_t+{\bf v}_{a}\cdot{\bm \nabla}$, and $\delta n_{b}$ follows analogously by exchanging indices $a\leftrightarrow b$ in Eq.~(\ref{soldeltan}).
This equation establishes a connection between the density and phase fluctuations, which facilitates experimental observation of the phase fields, as we will discuss later.
Substituting the expressions of $\delta n_{a(b)}$ into Eq.~(\ref{densityflu}) leads to the linear EoM for the phase fields alone:
\begin{equation}
\begin{aligned}
(\alpha\tilde{\nabla}_{a}^{2}-c_{a}^{2}\nabla^{2})\delta\theta_{a}&=\beta_{a}\tilde{\nabla}_{a}^{2}\delta\theta_{b},\\
(\alpha\tilde{\nabla}_{b}^{2}-c_{b}^{2}\nabla^{2})\delta\theta_{b}&=\beta_{b}\tilde{\nabla}_{b}^{2}\delta\theta_{a},
\end{aligned}
\label{phaseflu2}
\end{equation}
where $c_{a(b)}=\sqrt{U_{aa(bb)}\bar{n}_{a(b)}/m}$ is the sound velocity of a scalor BECs in the component $a(b)$. In Eq.~(\ref{phaseflu2}), $\alpha=U_{aa}U_{bb}/(U_{aa}U_{bb}-U_{ab}^{2})$ and $\beta_{a(b)}=U_{aa(bb)}U_{ab}/(U_{aa}U_{bb}-U_{ab}^{2})$ are dimensionless parameters solely determined by atomic interactions. As $U_{ab} =0$, $\alpha=1$ and $\beta_{a(b)}=0$, and two fields are decoupled.

In the following, we consider the case of $U_{aa}=U_{bb}\equiv U$, which is {a good approximation for the most commonly used alkaline atoms with spinor hyperfine states {\cite{hyperfine_states}}, 
 such as $^{87}$Rb {\cite{87Rb1,87Rb2}} and $^{23}$Na {\cite{23Na}}.} Moreover, we consider the two components have equal particle numbers and are in the mixed phase ($0<|U_{ab}|<U$) \cite{Uab<U1,Uab<U2,Uab<U3}, which leads to $\bar{n}_{a}=\bar{n}_{b}\equiv\bar{n}$ and $\bar{\theta}_{a}=\bar{\theta}_{b}$.
In this case, $\beta_{a}=\beta_{b}\equiv\beta$,  ${\bf v}_{a}={\bf v}_{b}\equiv{\bf v}_{\rm s}$ and $c_{a}=c_{b}\equiv c$. Then, Eq.~(\ref{phaseflu2}) can be decoupled into
\begin{equation}
\left[\left(\alpha\mp\beta\right)\tilde{\nabla}^{2}-c^{2}\nabla^{2}\right]\delta\theta_{\pm}=0,
\label{relphaseflu}
\end{equation}
where $\delta\theta_{+}=\delta\theta_{a}+\delta\theta_{b}$ and $\delta\theta_{-}=\delta\theta_{a}-\delta\theta_{b}$ are the re-defined fields in relation to the total and relative phase fluctuations, respectively. 

Equation~(\ref{relphaseflu}) describes two scalar fields in different curved (2+1)D spacetimes. According to the minimum coupling principle, the corresponding contra-covariant metrics 
are written as \cite{SM}
\begin{align}
g^{\mu\nu}_{\pm}=\left(\begin{array}{ccc}-1 & v_{\rm s} & v_{\rm s} \\v_{\rm s} & \frac{c^{2}}{\alpha\mp\beta}-v_{\rm s}^{2} & -v_{\rm s}^{2} \\v_{\rm s} & -v_{\rm s}^{2} & \frac{c^{2}}{\alpha\mp\beta}-v_{\rm s}^{2}\end{array}\right).
\label{metric}
\end{align}
The covariant metric $g_{\pm,\mu\nu}$ is the inverse of $g^{\mu\nu}_{\pm}$. 
Therefore, we end up with a dual phonon universe based on two relativistic phonon modes, respectively termed as the total phase phonon (TPP) and relative phase phonon (RPP) for brevity. These two modes experience different spacetime metrics indicated by $g^{\mu\nu}_{\pm}$, in which $\beta$ (i.e., $U_{ab}$) plays a key role. When $\beta=0$, the spacetime metrics of $\delta\theta_{\pm}$ become identical.

{\it Expanding/Contracting Dynamics-.} 
A prototype spacetime for studying the expansion or contraction of the universe is the Friedmann-Robertson-Walker (FRW) spacetime \cite{FRW1,FRW2,FRW3,FRW4}. 
By making $\beta$ (or equivalently $U_{ab}$) in Eq.~(\ref{metric}) time-dependent, we can derive the FRW metric.
To illustrate, we first consider a uniform BEC with $V=0$, such that $v_{\rm s}=0$ and $\bar{n}$ being a space-independent constant. 
Then, the line element is given by
\begin{align}
\label{dspm}
ds_{\pm}^{2}=-dt^{2}+a^{2}_{\pm}(t) d{\bf x}^{2},
\end{align}
where 
\begin{equation}
a^{2}_{\pm}(t)=\frac{m(U\mp U_{ab}(t))}{\bar{n}(U^{2}-U_{ab}^{2}(t))}
\label{scalefactor}
\end{equation}
is the scale factor, and $c=\sqrt{U\bar{n}/m}$ is the uniform sound velocity. 
Eq.~(\ref{dspm}) represents a homogeneous and isotropic FRW spacetime with flat 2D spatial sections (flat FRW). The time-dependent scale factor $a_{\pm}(t)$, which is achievable by tuning the magnetic field $B(t)$ near a Feshbach resonance in cold atoms, induces curvature in the (2+1)D spacetime.

\begin{figure}[t]
	\centering
	\includegraphics[width=0.45\textwidth]{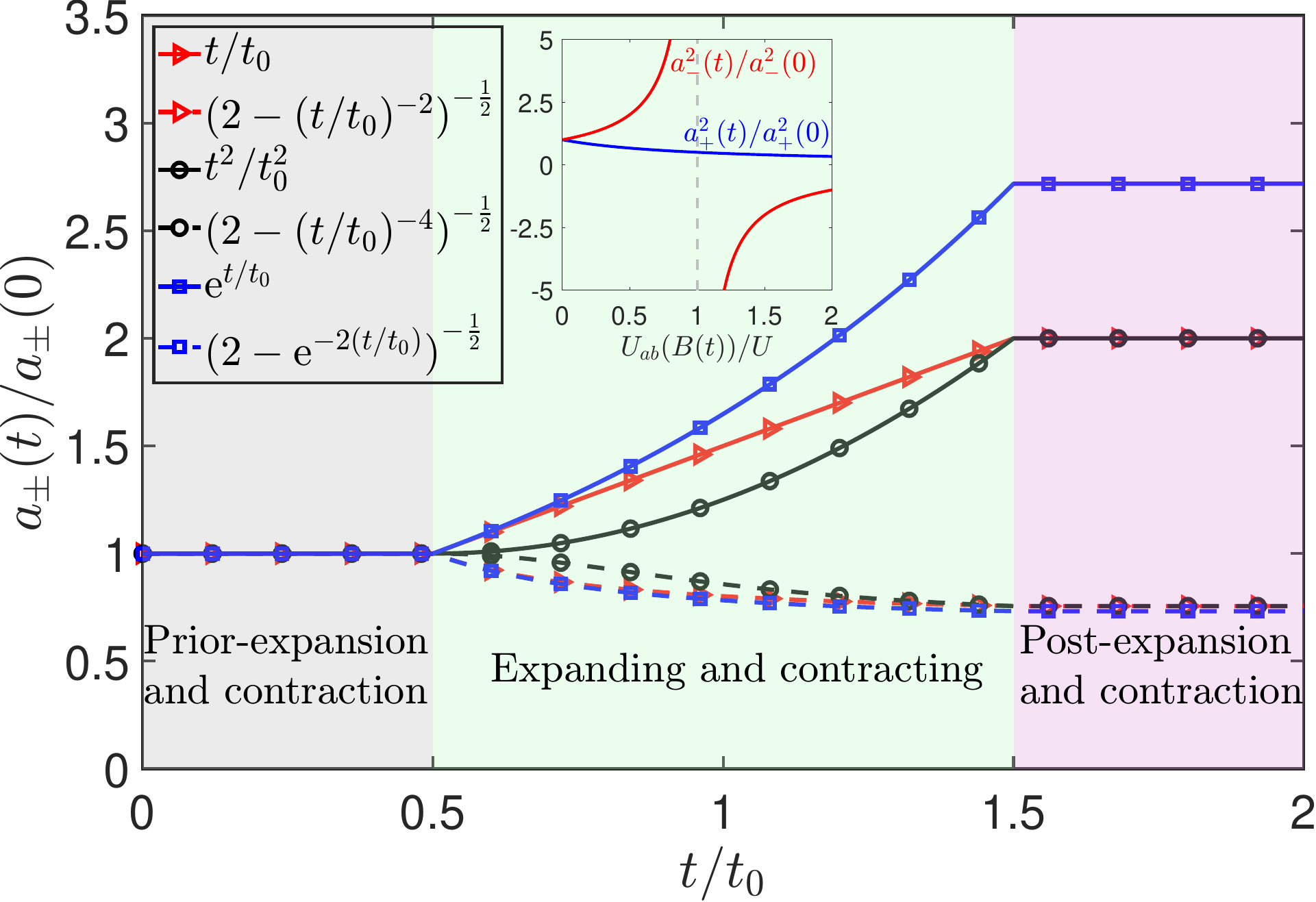}
	\caption{Expansion and contraction of the dual universe. Main panel: The dynamics of scale factors $a_{\pm}(t)$, with $t_{0}=\hbar/(U\bar{n}_{0})$ being the time unit. A solid line and the corresponding dashed line (with the same color and markers) denote the TPP and RPP universes, respectively. Various line markers (with triangles, circles, and squares) correspond to various ramping processes of $U_{ab}(t) < U$. 
  Inset: A case of ramping $U_{ab}$ across the BEC's phase transition point $U_{ab}=U$, at which $a_-$ diverges. Note that, the mixed BEC would be unstable in the regime $U_{ab}>U$.
  }
	\label{Fig2}
\end{figure}

According to Eq.~(\ref{scalefactor}), as $U_{ab}$ varies, the tendencies of $a_{+}(t)$ and $a_{-}(t)$ are always opposite. 
It means that one universe expands, and in the meanwhile, the other universe contracts. This serves as one of the key results of this paper.
Furthermore, by engineering the ramping behavior of $B(t)$, one can achieve arbitrary expasion/contraction processes $a_{\pm}(t)$. 
In Fig.~\ref{Fig2}, we show the tendencies of $a_{\pm}(t)$ when $U_{ab}$ ramps up in the manners of $U_{a b}(\tilde{t})/{U} \propto \tilde{t}^{-2}-1, \tilde{t}^{-4}-1$, and $e^{-2 \tilde{t}}-1$, where $\tilde{t}=t/t_{0}$ and $t_{0}=\hbar/(U\bar{n}_{0})$. These cases respectively leads to the linear (solid line with triangles), quadratic (with circles), and exponential (with squares) expansion processes and the corresponding contractions (dashed lines). All ramping processes start from $t_{\rm i} = 0.5 t_0$ and end at $t_{\rm f} = 1.5 t_0$. 
It is clearly indicated that while the detailed behavior of $a_{\pm}(t)$ varies with different ramping methods, the overall trends remain consistent: the TPP universe with $a_{+}(t)$ contracts, whereas the RPP universe with $a_{-}(t)$ expands.

We additionally noted that, according to Eq.~(\ref{scalefactor}), the factors $a^2_\pm$ turns to be negative if $U_{ab}>U$, i.e., $a_\pm$ are imaginary. This is understandable since pseudospin-1/2 BECs undergo a phase transition from mixed phase to phase separation at $U_{ab}=U$ \cite{Uab<U1,Uab<U2,Uab<U3}, at which $a_-^2$ turns to diverge, as illustrated in the inset of Fig.~\ref{Fig2}. Our discussions are only applicable to the mixed phase. Hence, in the main panel of Fig.~\ref{Fig2}, all ramping curves of $U_{ab}$ are properly scaled to ensure $U_{ab}(t)<U$.

For a non-uniform BEC, we can realize a FRW spacetime with curved spatial sections (curved FRW). For example, we consider central-force trapping potentials such that the density distributions are isotropic and varying along the radial direction $\rho$, i.e., $\bar{n}^{(\prime)}=\bar{n}_{0}\left(1-f(\rho)/R^{2}\right)$ and $f(\rho)=\pm 2\rho^{2}-\rho^{4}/R^{2}$, where $\bar{n}$ and $\bar{n}^{(\prime)}$ respectively correspond to the cases with positive and negative signs in $f(\rho)$, and $R$ is a control parameter with the same dimension as $\rho$. 
After a coordinate transformation $w=\rho(1\mp\rho^{2}/R^{2})^{-1}$, the line element is given by
\begin{equation}
ds_{\pm}^{2}=-dt^{2}+a^{2}_{\pm}(t) \left(\frac{dw^{2}}{1-\kappa w^{2}}+w^{2}d\varphi^{2}\right),
\label{dspm2}
\end{equation}
where $\kappa=\mp4/R^{2}$ is the spatial curvature. As $R\to\infty$, $\kappa\to0$ such that the curved FRW reduces to the flat one. 
Eq.~(\ref{dspm2}) indicates that, for the two different density distributions $\bar{n}$ and $\bar{n}^{(\prime)}$, the curved FRW can exhibit positive and negative spatial curvatures in the spacetime $(t,w,\varphi)$. However, for a given density distribution, the spatial curvature for both the TPP and RPP universes are the same.
Furthermore, based on the same ramping protocol as shown in Fig.~\ref{Fig2}, one can also realize the expansion and contraction of the curved FRW.

{\it Particles' Production-.} 
The particles' production during the expanding/contracting process 
may provide insights into the formation of elementary particles in the early Big Bang.
According to Eq.~(\ref{dspm}), the EoM of 
$\delta\theta_{\pm}$ is given by
\begin{equation}
\partial_{t}^{2}\delta\theta_{\pm}+\frac{2\dot{a}_{\pm}(t)}{a_{\pm}(t)}\partial_{t}\delta\theta_{\pm}-\frac{1}{a^{2}_{\pm}(t)}\nabla^{2}\delta\theta_{\pm}=0,
\label{expandingFRW}
\end{equation}
where $\dot{a}_{\pm}$ denotes the time-derivative of $a_{\pm}$ \cite{SM}.
For the curved FRW, one needs to replace the Laplace operator $\nabla^{2}$ by the Laplace-Beltrami operator. 
For a slow expanding/contracting, i.e., $\dot{a}(t)\approx0$, Eq.~(\ref{expandingFRW}) reduces to the Klein-Gordon equation. 

Taking the flat FRW as an example, we illustrate the particle generation \cite{particle2,particle1,Particle3}. We first quantize the fields $\delta\theta_{\pm}$ as
\begin{align}
\delta{\hat \theta}_{\pm}=\int\frac{kdk}{2\pi} \sum_{m}&\left[\hat{b}_{km}J_{m}(k\rho)e^{im\varphi}v_{\pm,k}(t)\nonumber\right.\nonumber\\
&\left.+\hat{b}^{\dagger}_{km}J_{m}(k\rho)e^{-im\varphi}v^{*}_{\pm,k}(t)\right],
\end{align}
where $J_{m}(\cdot)$ denotes the Bessel function of the first kind, $m=0,\pm1,\pm2,\cdots$ is the quantum number of the angular momentum $\hat L_{z}$, $k$ is the wavenumber, and $\hat{b}_{km}$ is the annihilation operator of the mode $v_{\pm,k}(t)$, satisfying bosonic commutation relations $[\hat{b}_{km},\hat{b}_{k'm'}^{\dagger}]=2\pi\delta_{mm'}\delta(k-k')/k$. The mode equation is given by
\begin{align}
\label{modeequ}
\partial_{t}^{2}v_{\pm,k}(t)+\frac{2\dot{a}_{\pm}(t)}{a_{\pm}(t)}\partial_{t}v_{\pm,k}(t)+\frac{f(k)}{a^{2}_{\pm}(t)}v_{\pm,k}(t)=0,
\end{align}
where $f(k)=k^{2}$. For the 
curved FRW, $f(k)$ will be $\kappa$-dependent \cite{Curved1,curve2,curve3,S_k3}. 

Since the scale factors $a_{\pm}(t)$ are time-independent before and after the ramping process (see Fig.~\ref{Fig2}), Eq.~(\ref{modeequ}) can be reduced to the EoM of simple harmonic oscillators.
Then, the linear-independent solutions are worked out as
\begin{equation}
\label{vksolution}
v_{\pm,k}=
\begin{cases}
\frac{{e}^{-i\omega^{\rm i}_{\pm,k}t}}{\sqrt{2\hbar ka_{\pm}(t_{\rm i})}} & t<t_{\rm i} \\
\tilde{v}_{\pm,k}(t) & t_{\rm i}\le t<t_{\rm f} \\
\frac{S{e}^{-i\omega^{\rm f}_{\pm,k}t}+T{e}^{i\omega^{\rm f}_{\pm,k}t}}{\sqrt{2\hbar ka_{\pm}(t_{\rm f})}} & t\ge t_{\rm f}
\end{cases}
\end{equation}
and
\begin{equation}
\label{uksolution}
u_{\pm,k}=\begin{cases}
\frac{S^{*}{e}^{-i\omega^{\rm i}_{\pm,k}t}-T{e}^{i\omega^{\rm i}_{\pm,k}t}}{\sqrt{2\hbar ka_{\pm}(t_{\rm i})}} & t<t_{\rm i} \\
\tilde{u}_{\pm,k}(t) & t_{\rm i}\le t<t_{\rm f} \\
\frac{{e}^{-i\omega^{\rm f}_{\pm,k}t}}{\sqrt{2\hbar ka_{\pm}(t_{\rm f})}} & t\ge t_{\rm f}
\end{cases}
\end{equation}
where $t_{\rm i}$ and $t_{\rm f}$ are the initial and final time of expansion, $\omega_{\pm,k}^{\rm i(f)}=k/a_{\pm}(t_{\rm i(f)})$ is the initial (final) frequency. $S$ and $T$ are coefficients to be determined by the connecting condition at $t_{\rm i}$ and $t_{\rm f}$. $\tilde{v}_{\pm,k}(t)$ and $\tilde{u}_{\pm,k}(t)$ can be obtained by numerically solving the mode equation in Eq.~(\ref{modeequ}). 
The normalization condition for the mode function is ${\rm Wr}[v_{\pm, k}(t),v^{*}_{\pm, k}(t)]=i$ where ${\rm Wr}[\cdot\ ,\cdot]$ represents the Wronskian determinant. We define $\hat{b}'_{km}$ as the annihilation operator of phonon in the mode $u_{\pm,k}(t)$, and it also satisfies $[\hat{b}_{km}',(\hat{b}_{k'm'}')^{\dagger}]=2\pi\delta_{mm'}\delta(k-k')/k$. Notice that there is only positive frequency for $v_{\pm,k}(t<t_{\rm i})$ and $u_{\pm,k}(t>t_{\rm f})$. This means that the vacuum states for $t<t_{\rm i}$ and $t>t_{\rm f}$ are defined as $\hat{b}_{km}|\Omega_{\rm i}\rangle=0$ and $\hat{b}_{km}'|\Omega_{\rm f}\rangle=0$, respectively. Comparing Eq.~(\ref{vksolution}) and (\ref{uksolution}), we find $u_{\pm,k}(t)=S_{\pm}^{*}v_{\pm,k}(t)-T_{\pm}v^{*}_{\pm,k}(t)$ and $v_{\pm,k}(t)=S_{\pm}u_{\pm,k}(t)+T_{\pm}u^{*}_{\pm,k}(t)$, which satisfies $|S_{\pm}|^{2}-|T_{\pm}|^{2}=1$. 

There exists an intuitive understanding from the perspective of the 1D scattering problem in the time dimension~\cite{scattering1,scattering2}. When $t<t_{\rm i}$, $v_{\pm,k}(t)$ represents a plane wave towards to scattering potential. When $t>t_{\rm f}$, $v_{\pm,k}(t)$ is a linear combination of outgoing and incoming plane waves, and the respective probabilities are $|S_{\pm}|^{2}$ and $|T_{\pm}|^{2}$. The probability conservation requires $1+|T_{\pm}|^{2}=|S_{\pm}|^{2}$. The similar analysis can be applies to $u_{\pm,k}(t)$. Due to $u_{\pm,k}(t)$ and $v_{\pm,k}(t)$ are different modes of the same fields $\delta{\hat \theta}_{\pm}$, the resulting relation for the corresponding operator is
\begin{eqnarray}
\begin{aligned}
\label{bogoliubov}
\hat{b}_{km} &=S_{\pm}^{*}\hat{b}_{km}'-T_{\pm}^{*}{\hat b}'{}^{\dagger}_{k,-m};\\
\hat{b}_{km}'&=S_{\pm}\hat{b}_{km}-T_{\pm}^{*}{\hat b}_{k,-m}^{\dagger},
\end{aligned}
\end{eqnarray}
which is the Bogoliubov transformation of bosonic operators. Therefore, after the expansion/contraction, the particle-number expectation is $\langle\Omega_{\rm f}|\hat{b}^{\dagger}_{km}\hat{b}_{km}|\Omega_{\rm f}\rangle=|T|^{2}\neq0$, indicating the phonon appears during the dynamics. 

\begin{figure}[t]
	\centering
	\includegraphics[width=0.42\textwidth]{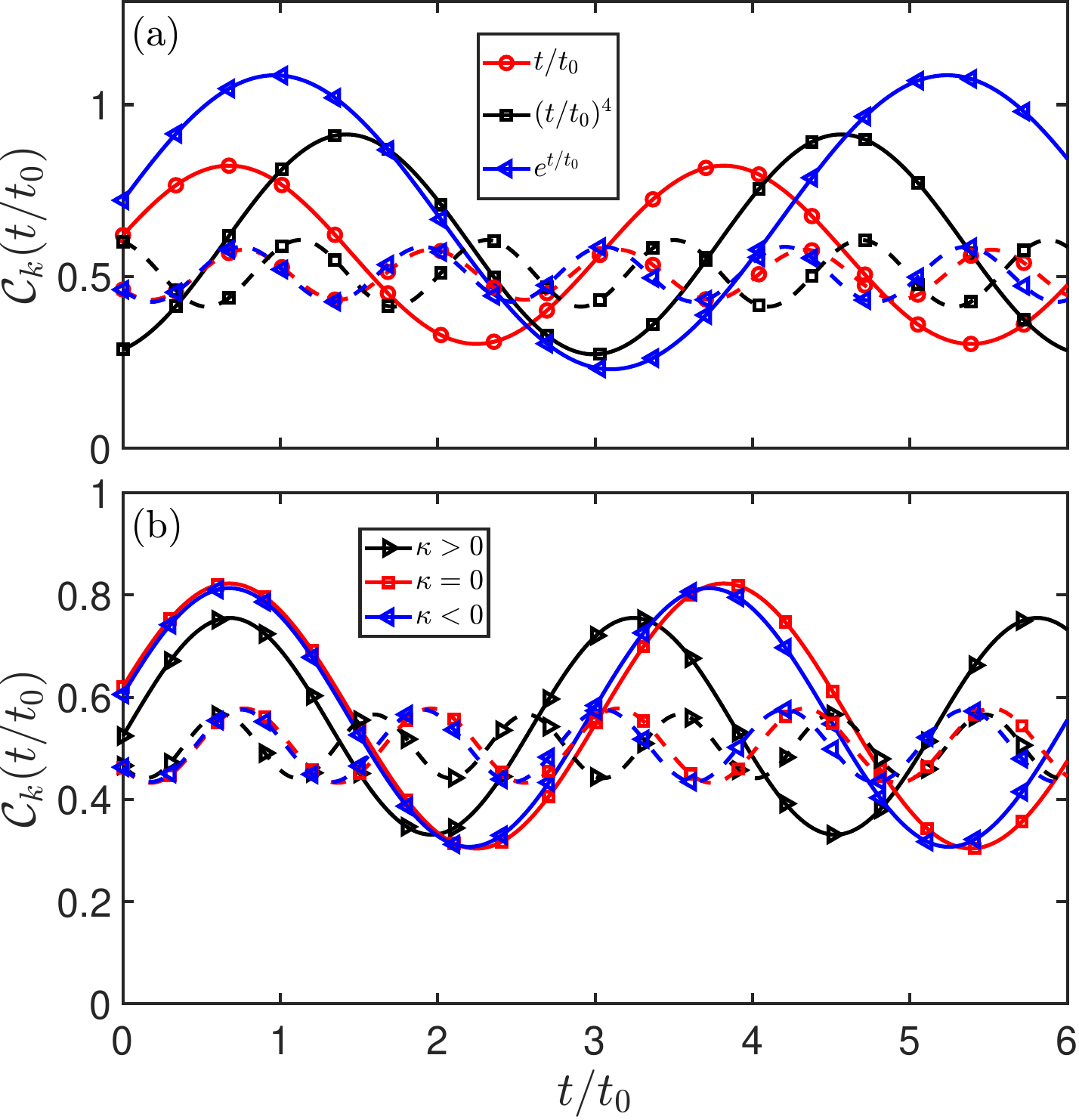}
	\caption{Correlator spectra of the SDC for the flat FRW (a) and the curved FRW (b). In panel (a), lines with different markers correspond to various ramping processes. In panel (b), lines with different markers correspond to various spatial curvatures $\kappa$.}
	\label{Fig3}
\end{figure}

{\it Experimental Observation-.} 
As mentioned earlier, the phase and density fluctuations are mutually related by Eq.~(\ref{soldeltan}). Hence, the universe expansion/contraction can be experimentally verified by detecting the density response, a quantity that is readily measurable in current experimental setups \cite{S_k1,S_k2,S_k3,Curved1,Curved2}. Specifically, we define the scaled density contrast (SDC)  for the total and relative fields as
\begin{align}
\label{SDC}
d_{\pm}(t,\rho)=\sqrt{\frac{\bar{n}(\rho)}{\bar{n}_{0}^{3}}}\left[\delta n_{\pm}(t,\rho)-\delta n_{\pm}(0,0)\right],
\end{align}
where $\delta n_{\pm}$ are the combinations of the measured bare fields $\delta n_{a,b}$. 
The equal-time correlator of SDC is defined as $\langle d_{\pm}(t,\rho)d_{\pm}(t,\rho')\rangle$, whose Fourier spectrum is 
\begin{equation}
{\cal C}_{\pm,k}(t)=\frac{1}{2}+|T_{\pm}|^{2}+|S_{\pm}T_{\pm}|\cos(\omega^{\rm f}_{\pm,k}t+\theta_{\pm,k}).
\label{Ck}
\end{equation}
The frequency and amplitude of ${\cal C}_{\pm,k}(t)$ 
encapsulate
the information of universe dynamics. 
Specifically, the oscillating frequencies $\omega^{\rm f}_{\pm,k} = k/a_{\pm}(t_{\rm f})$ are inversely proportional to the scale factors, while the amplitudes are determined by the expansion/contraction history through the coefficients $S_{\pm}$ and $T_{\pm}$.

In Fig.~\ref{Fig3}(a), we present ${\cal C}_{\pm,k}(t)$ of the flat FRW for different ramping processes 
illustrated in Fig.~\ref{Fig2}.
Clearly, the oscillation frequencies (solid curves) of the expanding universe are larger than those of the contracting one. This result is robust to the various expansion and contraction histories. Additionally, the amplitudes of ${\cal C}_{\pm,k}(t)$ exhibit variations depending on the ramping details.

The measurement for a curved FRW universe ($\kappa\neq0$) can be performed in a very similar way. 
In Fig.~\ref{Fig3}(b), we show ${\cal C}_{\pm,k}(t)$ for cases with various spatial curvatures $\kappa$, in which $\kappa = 0$ (lines with squares) denotes the flat FRW and the positive/negative $\kappa$ are respectively realized by using the aforementioned $\bar{n}$ and $\bar{n}^{(\prime)}$. The results show that the nonzero $\kappa$ leads to different oscillation frequencies and amplitudes, both of which are suppressed compared to the flat case.

{\it Summary-.} 
We have demonstrated that a pseudospin-1/2 BEC can simulate a dual phonon 
universe, where the total-phase and relative-phase phononic modes live in 
distinct spacetimes. By controlling the interspecies interaction, we can realize the simultaneous expansion of one universe and contraction of the other, accompanied by the creation of bosonic particles. These findings are proposed to be experimentally detected via the correlators of the density contrast. 
Future studies may be extended to spinor BECs with higher spins, where increased internal degrees of freedom may facilitate the simulation of more complex cosmological models and unlock richer multiverse-related physics.

\begin{acknowledgments}
We are grateful to Chenwei Lv and Yangqian Yan for stimulating discussions.
This work is supported by NSFC (Grants No.12174300, No.12374248, and No.12174236) and the Innovation Program for Quantum Science and Technology (Grants No.2021ZD0302001), the Fundamental Research Funds for the Central Universities (Grant No. 71211819000001) and Tang Scholar.
\end{acknowledgments}


%

\appendix
\begin{widetext}
\section{Supplemental Material of ``Dual Phonon Universe in a Spinor Bose Gas''} 
In this Supplemental Material, we provide the detailed derivation for: (1) the hydrodynamics equation of interacting Bose gases; (2) the equation of motion (EoM) of the density and phase fluctuation fields in Bose gases; (3) the metric of the phonon universe, and (4) the EoM of phonon in the Friedmann-Robertson-Walker (FRW) spacetime.

\subsection{1. Hydrodynamics equation}
The Gross-Pitaevskii equation [Eq.~(2) of the main text] can be derived by minimizing the action $S$ [Eq.~(1) of the main text] with respect to $\phi_{a(b)}$. Specifically, by taking the polar representation of the bosonic fields
\begin{equation}
    \phi_{a(b)}=\sqrt{n_{a(b)}}e^{i\theta_{a(b)}},
    \label{polar}
    \tag{S1}
\end{equation}
the action can be recast into
\begin{equation}
{\cal S}=\sum_{i=a,b}\int dt d^2{\bf x} \left[
i\hbar\left(\frac{1}{2}\partial_t n_i+in_i\partial_t \theta_i\right)
-\frac{\hbar^2}{2m}\left[{\bm \nabla}\left(\sqrt{n}_ie^{-i\theta_i}\right)\right]\cdot\left[{\bm \nabla}\left(\sqrt{n_i}e^{i\theta_i}\right)\right]\right]-\left[Vn_i-\frac{U_{ii}}{2}n_i^2-U_{ab}n_a n_b\right].
\label{action}
\tag{S2}
\end{equation}
We expand the second term as
\begin{equation}
  \begin{aligned}
    \left[{\bm \nabla}\left(\sqrt{n}_ie^{-i\theta_i}\right)\right]\cdot\left[{\bm \nabla}\left(\sqrt{n_i}e^{i\theta_i}\right)\right]&=\left(\frac{1}{2\sqrt{n_i}}({\bm \nabla} n_i)e^{-i\theta_i}-i\sqrt{n_i}({\bm \nabla} \theta_i)e^{-i\theta_i}\right)\cdot\left(\frac{1}{2\sqrt{n_i}}({\bm \nabla} n_i)e^{i\theta_i}+i\sqrt{n_i}({\bm \nabla} \theta_i)e^{i\theta_i}\right)\\
    &=\left[\frac{1}{4n_i}({\bm \nabla} n_i)\cdot({\bm \nabla} n_i)+n_i(\bm\nabla\theta_i)(\bm\nabla\theta_i)\right].
  \end{aligned}
  \label{action2}
  \tag{S3}
\end{equation}
Then, the action can be rewritten as
\begin{equation}
     \begin{aligned}
{\cal S}=\sum_{i=a,b}\int dt d^2{\bf x} \left[
i\hbar\left(\frac{1}{2}\partial_t n_i+in_i\partial_t \theta_i\right)
-\frac{\hbar^2}{2m}\left[\frac{1}{4n_i}({\bm \nabla} n_i)\cdot({\bm \nabla} n_i)+n_i({\bm \nabla}\theta_i)\cdot({\bm \nabla}\theta_i)\right]
-Vn_i-\frac{U_{ii}}{2}n_i^2-U_{ab}n_a n_b\right].
    \end{aligned}
    \label{action3}
    \tag{S4}
\end{equation}
The stationary point of ${\cal S}$ with respect to $n_{i}$ leads to the EoM of $n_{i}$, 
\begin{equation}
\label{newton}
    \hbar\partial_{t}\theta_{i}=\frac{\hbar^{2}}{2m}\left[\frac{\nabla^{2} n_{i}}{4n^2_{i}}+\frac{1}{2}{\bm \nabla}\cdot\left(n_{i}^{-1}{\bm \nabla} n_{i}^{-1}\right)-\nabla^{2}\theta_{i}\right]-V
-U_{ii}n_{i}-U_{ab}n_{j},\quad i,j=(a,b),i\neq j.
\tag{S5}
\end{equation}
Similarly, 
the stationary point of ${\cal S}$ with respect to $\theta_{i}$ leads to the 
equation for continuity
\begin{equation}
\label{continuity}
    \partial_{t}n_{i}+\frac{\hbar}{m}{\bm \nabla}\cdot(n_{i}{\bm \nabla} \theta_{i})=0,\quad i=(a,b).
    \tag{S6}
\end{equation}
Equations (\ref{newton}) and (\ref{continuity}) reproduce the hydrodynamics equations (2) and (3) in the main text.

\subsection{2. Equation of motion of the fluctuation-fields}
We denote the steady solution to the hydrodynamics equation as $\overline{n}_i$ and $\overline{\theta}_i$, and the respective fluctuation as $\delta n_i$ and $\delta\theta_i$. Therefore, we have
\begin{equation}
\label{densityandphase}
    n_i=\overline{n}_i+\delta n_i; \quad \theta_i=\overline{\theta}_i+\delta\theta_i.
    \tag{S7}
\end{equation}
By substituting Eq.~(\ref{densityandphase}) into Eq.~(\ref{newton}) and (\ref{continuity}), we have
\begin{equation}
\begin{aligned}
\label{9}
    \hbar\partial_{t}(\overline{\theta}_{i}+\delta\theta_i)
    &=\frac{\hbar^{2}}{2m}\left[\frac{{\nabla}^{2} (\overline{n}_{i}+\delta n_i)}{4(\overline{n}_i+\delta n_i)^2}+\frac{1}{2}{ {\bm\nabla}}\cdot\left[(\overline{n}_{i}+\delta n_i)^{-1}{ {\bm\nabla}} (\overline{n}_{i}+\delta n_i)^{-1}\right]-{\nabla}^{2}(\overline{\theta}_{i}+\delta\theta_i)\right]
\\&-V-U_{ii}(\overline{n}_{i}+\delta n_i)-U_{ab}(\overline{n}_{j}+\delta n_j);
\end{aligned}
\tag{S8}
\end{equation}
and
\begin{equation}
\begin{aligned}
&\partial_{t}(\overline{n}_{i}+\delta n_i)+\frac{\hbar}{m}{ \bm\nabla}\cdot[(\overline{n}_{i}+\delta n_i){ \bm\nabla} (\overline{\theta}_{i}+\delta\theta_i)]=0.
\end{aligned}
\tag{S9}
\end{equation}
Considering stationary part has solved the hydrodynamics equation, we end up with the equations of the fluctuation part. The time derivative of $\delta \theta_{i}$ is written as
\begin{equation}
\begin{aligned}
    &\hbar\partial_{t}\delta \theta_{i}\\&=\frac{\hbar^2}{2m}\left[\frac{1}{2\overline{n}_i^2}\left(({\bm\nabla\overline{n}_i})({\bm\nabla\delta n_i})-\frac{\delta n_i}{\overline{n}_i}{\bm\nabla}^2\overline{n}_i\right)+\frac{1}{2}{\bm\nabla}\left(\frac{1}{\overline{n}_i}{\bm\nabla}\delta n_i-\frac{\delta n_i}{\overline{n}_i}{\bm\nabla}\overline{n}_i\right)-2({\bm\nabla\overline{\theta}_i})({\bm\nabla\delta\theta_i})\right]-U_{ii}\delta n_{i}-U_{ab}\delta n_{j}
    \end{aligned}
    \tag{S10}
\end{equation}
For uniform Bose gases, the second spatial derivative of density and density fluctuation can be ignored. 
Therefore, we obtained the EoM for the fluctuation fields,
\begin{align}
\hbar\partial_{t}\delta \theta_{i}&=-\hbar\cdot{\bf v}_i({\bm\nabla\delta\theta_i})-U_{ii}\delta n_{i}-U_{ab}\delta n_{j}; \tag{S11}\\
    \partial_{t}\delta n_{i}&=-{\bm\nabla}\cdot\left[(\delta n_{i}){\bf v}_{i}+\frac{\hbar}{m}\bar{n}_{i}{\bm \nabla} \delta\theta_{i}\right].
    \tag{S12}
\end{align}
which reproduce the Eqs.~(4) and (5) in the main text, with ${\bf v}_{i}=(\hbar{\bm \nabla}{\bar \theta}_{i})/m$ being the superfluid velocity. 
\subsection{3. Metric of the phonon universe}
It can be verified that the EoM [i.e., Eq.~(8) of the main text]
\begin{equation}
    \left[\left(\alpha\mp\beta\right)(\partial_t+ {\bf v}_i\cdot{\bm \nabla} )^{2}-c^{2}\nabla^{2}\right]\delta\theta_{\pm}=0,
    \tag{S13}
\end{equation}
can be obtained by minimizing the action
\begin{equation}
    \begin{aligned}
    {\cal S}_{\delta\theta_{\pm}}&=\frac{{\hbar}^2}{2}\int dtd^2{\bf x}\left[\left(\alpha\mp\beta\right)((\partial_t+ {\bf v}_i\cdot{\bm\nabla} ){\delta\theta_{\pm}})^2-c^2\nabla^2\delta\theta_{\pm}\right]
    \end{aligned}
    \tag{S14}
\end{equation}
with respect to the field $\delta\theta_{\pm}(t,x,y)$.
In order to express ${\cal S}_{\delta\theta_{\pm}}$ as a action of scalar field in a curved spacetime, we expand the integrand as
\begin{equation}
    \begin{aligned}
 &\left[\left(\alpha\mp\beta\right)((\partial_t+ {\bf v}_i\cdot{\bm\nabla} ){\delta\theta_{\pm}})^2-c^2({\bm\nabla}\delta\theta_{\pm})({\bm\nabla}\delta\theta_{\pm})\right]\\
&=\left[(\alpha\mp\beta)\left[(\partial_t^2\delta\theta_{\pm})+v_i^2({\bm\nabla}\delta\theta_{\pm})({\bm\nabla}\delta\theta_{\pm})+{\bf v}_i(\partial_t\delta\theta_{\pm})({\bm\nabla}\delta\theta_{\pm})+{\bf v}_i(\partial_t\delta\theta_{\pm})({\bm\nabla}\delta\theta_{\pm})\right]-c^2({\bm\nabla}\delta\theta_{\pm})({\bm\nabla}\delta\theta_{\pm})\right]\\
    &=\left(\begin{array}{ccc}\partial_t\delta\theta_{\pm} & {\bm\nabla}\delta\theta_{\pm}  \end{array}\right)\left(\begin{array}{ccc}(\alpha\mp\beta) & (\alpha\mp\beta){\bf v}_i  \\(\alpha\mp\beta){\bf v}_i & v_i^2(\alpha\mp\beta)-c^2\delta_{xy} & \end{array}\right)\left(\begin{array}{ccc}\partial_t\delta\theta_{\pm} \\ {\bm\nabla}\delta\theta_{\pm}  \end{array}\right)\\
    &=\left(\begin{array}{ccc}\partial_t\delta\theta_{\pm} & {\bm\nabla}\delta\theta_{\pm}  \end{array}\right)\mathcal{M}\left(\begin{array}{ccc}\partial_t\delta\theta_{\pm} \\ {\bm\nabla}\delta\theta_{\pm}  \end{array}\right).
    \end{aligned}
    \label{action4}
    \tag{S15}
\end{equation}
Therefore, the action can also be written in covariant form 
\begin{equation}
\begin{aligned}
    {\cal S}_{\delta\theta_{\pm}}&=-\frac{{\hbar}^2}{2}\int dtd^2{\bf x}\sqrt{|\det(g_{\pm,\mu\nu})|}g^{\mu\nu}_{\pm}({\partial_{\mu}}\delta\theta_{\pm})({\partial_{\nu}}\delta\theta_{\pm})\\
       &=\frac{{\hbar}^2}{2}\int dtd^2{\bf x}\frac{-g^{\mu\nu}_{\pm}}{\sqrt{|\det(g^{\mu\nu}_{\pm})|}}({\partial_{\mu}}\delta\theta_{\pm})({\partial_{\nu}}\delta\theta_{\pm}).\\
\end{aligned}
\label{action5}
\tag{S16}
\end{equation}
By comparing Eq.~(\ref{action4}) and (\ref{action5}), we find
\begin{align}
\frac{-g^{\mu\nu}_{\pm}}{\sqrt{|\det(g^{\mu\nu}_{\pm})|}}=\mathcal{M}=\left(\begin{array}{ccc}(\alpha\mp\beta) & (\alpha\mp\beta){\bf v}_i  \\(\alpha\mp\beta){\bf v}_i & v^2_i(\alpha\mp\beta)-c^2\delta_{xy} & \end{array}\right).
\label{metric}
\tag{S17}
\end{align}
Here, we have assumed that the superfluid velocity is isotropic, i.e., ${\bf v}_i={\bf v}_s$. According to Eq. (\ref{metric}), we can derive the expression of the metric 
\begin{equation}
    g^{\mu\nu}_{\pm}=\frac{-\mathcal{M}}{\det\mathcal{M}}=\left(\frac{\alpha\mp\beta}{c^2}\right)^2\left(\begin{array}{ccc}-1 & {\bf v}_s  \\{\bf v}_s & \frac{c^2}{(\alpha\mp\beta)}\delta_{xy}-v^2_s & \end{array}\right).
    \tag{S18}
\end{equation}
For the (2+1)d spacetime, it can be simplified as
\begin{equation}
    g^{\mu\nu}_{\pm}=\left(\begin{array}{ccc}-1 & { v}_s &{ v}_s \\{ v}_s & \frac{c^2}{(\alpha\mp\beta)}-{v}_s^2 &-v_s^2 \\{ v}_s & -v_s^2 & \frac{c^2} 
 {(\alpha\mp\beta)}-v_s^2 \end{array}\right),
    \tag{S19}
\end{equation}
which yields Eq.~(9) of the main text.
When ${\bf v}_s=0$, we have
\begin{equation}
    g^{\mu\nu}_{\pm}=\left(\begin{array}{ccc}-1 & 0 & 0 \\ 0 & \frac{c^2}{(\alpha\mp\beta)} & 0 \\ 0 & 0 & \frac{c^2} 
 {(\alpha\mp\beta)} \end{array}\right),
    \tag{S20}
\end{equation}
and
\begin{equation}
    g_{\pm,\mu\nu}=\left(\begin{array}{ccc}-1 & 0 & 0 \\ 0 & \frac{(\alpha\mp\beta)}{c^2} & 0 \\ 0 & 0 &\frac{(\alpha\mp\beta)}{c^2}  \end{array}\right),
    \tag{S21}
\end{equation}
which yields $\sqrt{|g|}\equiv\sqrt{|\det(g^{\mu\nu}_{\pm})|}=(\alpha\mp\beta)/c^2$.

\subsection{4. Scalar field in FRW universe}
The action of the massless scalar field $\delta\theta_{\pm}(t,x,y)$ in curved spacetime can be written as 
\begin{equation}
    \begin{aligned}
    {\cal S}
    &=-\frac{\hbar^2}{2}\int dtdxdy\left[\sqrt{|g|}g^{\mu\nu}_{\pm}(\partial_{\mu}\delta\theta_{\pm})(\partial_{\nu}\delta\theta_{\pm})\right].
       \end{aligned}
    \tag{S22}
\end{equation}
According to the Euler-Lagrangian equation,  we have the EoM of the field
\begin{equation}
\begin{aligned}
\label{24}
    \partial_{\mu}(\sqrt{|g|}g^{\mu\nu}_{\pm}(\partial_{\nu}\delta\theta_{\pm}))=0
\end{aligned}
\tag{S23}
\end{equation}
For the (2+1)d FRW, which describes the universe expanding uniformly in all directions in the space part, the metric can be written as
\begin{equation}
\label{25}
    ds^2=g_{\pm,\mu\nu}dx^{\mu}dx^{\nu}=-dt^2+a^2(t)\left(dx^2+dy^2\right).
    \tag{S24}
\end{equation}
By substituting the specific form of the metric (\ref{25}) into Eq.~(\ref{24}), we finally obtain the Klein-Gordon equation of the massless scalar field $\delta\theta_{\pm}(t,x,y)$ in the curved spacetime 
\begin{equation}
\partial_t^2\delta\theta_{\pm}+2\frac{\dot{a}_{\pm}(t)}{a_{\pm}(t)}\partial_t\delta\theta_{\pm}-\frac{1}{a_{\pm}^2(t)}\nabla^2\delta\theta_{\pm}=0.
    \tag{S25}
\end{equation}
which reproduces the Eq.~(13) of the main text.
\end{widetext}
\end{document}